\RequirePackage{lineno}
\documentclass[twocolumn,showpacs,preprintnumbers,prb]{revtex4-1}
\usepackage{graphicx,bm,amsmath,amssymb}

\def\gz{\ifmmode{Z\hskip -4.8pt Z}
    \else{\hbox{$Z\hskip -4.8pt Z$}}\fi}

\newcommand{\be}{\begin{equation}}
\newcommand{\ee}{\end{equation}}
\newcommand{\bea}{\begin{eqnarray}}
\newcommand{\eea}{\end{eqnarray}}

\begin{document}

\title{Nonequilibrium self-energies, Ng approach and heat current of a nanodevice for small bias voltage and temperature}
\author{A.~A.~Aligia}
\affiliation{Centro At\'{o}mico Bariloche and Instituto Balseiro, Comisi\'{o}n Nacional
de Energ\'{\i}a At\'{o}mica, 8400 Bariloche, Argentina}
\email{aligia@cab.cnea.gov.ar}

\begin{abstract}
Using non-equilibrium renormalized perturbation theory to second order in
the renormalized Coulomb repulsion, we calculate the lesser $\Sigma^<$ and
and greater $\Sigma^>$ self-energies of the impurity Anderson model, which
describes the current through a quantum dot, in the general asymmetric case.
While in general a numerical integration is required to evaluate the
perturbative result, we derive an analytical approximation for small
frequency $\omega$, bias voltage $V$ and temperature $T$ which is exact to
total second order in these quantities. The approximation is valid when the
corresponding energies $\hbar \omega$, $eV$ and $k_B T$ are small compared
to $k_B T_K$, where $T_K$ is the Kondo temperature. The result of the
numerical integration is compared with the analytical one and with Ng
approximation, in which $\Sigma^<$ and $\Sigma^>$ are assumed proportional
to the retarded self-energy $\Sigma^r$ times an average Fermi function.
While it fails at $T=0$ for $\hbar|\omega | \lesssim eV$ we find that the Ng
approximation is excellent for $k_B T > eV/2$ and improves for asymmetric
coupling to the leads. Even at $T=0$, the effect of the Ng approximation on
the total occupation at the dot is very small. The dependence on $\omega$
and $V$ are discussed in comparison with a Ward identity that is fulfilled
by the three approaches. We also calculate the heat currents between the dot and any of the leads 
at finite bias voltage. One of the heat currents changes sign with the applied bias voltage 
at finite temperature.
\end{abstract}

\maketitle



\section{Introduction}

\label{intro}

Progress in nanotechnology has led to the confinement of electrons into
small regions, where the electron-electron interactions become increasingly
important. Therefore, the interpretation of transport experiments at finite
bias voltage $V$, for example different variants of the Kondo effect in
transport through quantum dots (QDs),\cite%
{gold,cro,wiel,grobis,parks,serge,ama,keller} requires the theoretical
treatment of the effects of both, nonequilibrium physics and strong
correlations. This problem is very hard and at present only approximate
treatments are used which have different limitations.\cite{none,hbo,rosch}

For nonequilibrium problems, perturbation theory is performed on the Keldysh
contour, in which the time evolves from $t_{0}\rightarrow -\infty $ in which
the system is in a well defined state and the perturbation is absent, to $%
t\rightarrow +\infty $ in one branch and returns to the initial state at $%
t_{0}$ on the other branch of the contour. Thus, the position in the contour
is not only given by the time, but also by a branch index. See for example
Ref. \onlinecite {lif} from which we borrow the notation. As a consequence,
there are four different one-particle Green functions depending on the
branch index of the creation and annihilation operators. They can be
classified \ as retarded, advanced, lesser and greater ($G^{r}$, $G^{a}$, $%
G^{<}$ and $G^{>}$ respectively). Similarly, the Dyson equation leads to
four self-energies $\Sigma ^{r}$, $\Sigma ^{a}$, $\Sigma ^{<}$ and $\Sigma
^{>}$.\cite{none,lif}

In general, it is more difficult to approximate the lesser and greater
quantities than the retarded ones. Ng proposed an approximation in which the
lesser and greater self-energies $\Sigma ^{<}$ and $\Sigma ^{>}$ are
proportional to average distribution functions ($\tilde{f}(\omega )$ and $%
\tilde{f}(\omega )-1$ respectively, see Section \ref{nga}) with the same
proportionality factor.\cite{ng} A consistency equation [Eq. (\ref{sdif})]
imposes that this factor is the imaginary part of the retarded self-energy $%
\Sigma ^{r}$. Therefore, this approximation permits to reduce the problem to
the calculation of retarded quantities only. The Ng approximation has been
used in many different subjects, like Andreev tunneling through strongly
interacting QDs,\cite{faz} spin polarized transport,\cite{pz,ser,zz} first
principles calculations of correlated transport through nanojunctions,\cite%
{fer} thermopower, \cite{dong} decoherence effects \cite{rap} and scaling 
\cite{bal} in transport through QDs, magnetotransport in graphene,\cite{ding}
asymmetric effects of the magnetic field in an Aharonov-Bohm interferometer,%
\cite{lim} and shot noise in QDs irradiated with microwave fields.\cite{zhao}
Therefore, it is of interest to test this approximation and establish its
range of validity. In a recent Letter,\cite{mun} is was claimed that Ng
approximation (Section \ref{nga}) is exact at low energies. In a Comment to
this work we have argued that it is not the case.\cite{com} In their Reply,%
\cite{reply} the authors claim that our analytical result for $\Sigma ^{<}$
for zero temperature derived previously does not satisfy a Ward identity,
but a direct calculation shows that it does.\cite{rr,note2} This will be
shown for all temperatures in Section \ref{ward}.

One of the approaches used to study the impurity Anderson model (IAM) out of
equilibrium is Keldysh perturbation theory in the Coulomb repulsion 
$U$.\cite{none,hersh,levy,fuji,hama} However, it is restricted to small values of $U$. 
Instead, in renormalized perturbation theory (RPT),\cite{he1} the
renormalized repulsion $\widetilde{U}$ is always small allowing for a
perturbation expansion even if $U\rightarrow \infty$. A calculation of $%
\Sigma^r$ to second order in $\widetilde{U}$ leads to the exact result to
total second order in frequency $\omega$, bias voltage $V$ and temperature $T
$ in terms of thermodynamic quantities, or the renormalized parameters which
can be obtained from exact Bethe ansatz \cite{ogu2} or
numerical-renormalization-group (NRG) \cite{re1,re2} calculations at
equilibrium. This $\Sigma^r$ has been used to obtain the exact form of the
conductance through a quantum dot to total second order in $V$ and $T$ for
the electron-hole symmetric (EHS) IAM with symmetric voltage drops and
coupling to the leads.\cite{ogu2} These results are valid for $eV$ and $k_B T$ 
small compared to $k_B T_K$, where $T_K$ is the Kondo temperature.
Motivated by recent experiments searching for universal scaling relations
for the conductance,\cite{grobis,scott}, further developments were 
made,\cite{bal,rinc,sela,roura,scali} but concentrated mainly on the EHS case.

Besides, thermal properties of quantum dots have been studied before,\cite{dong,boese,kim,vel,ss,see}
but concentrated mainly on the linear response regime of vanishing voltage and temperature
gradient.  

In this work we calculate the lesser and greater self-energies of the IAM in
the general (not EHS) case, for different (symmetric and asymmetric)
coupling to the leads, using RPT to second order in $\widetilde{U}$. The
result is compared with the Ng approximation for different temperatures. 
We derive an exact analytical expression for small $\omega $, $V$ and $T$
(to total second order), useful when the corresponding energies $\hbar
\omega $, $eV$ and $k_{B}T$ are small compared to $k_{B}T_{K}$.
We also calculate the heat currents between the dot and any of
the leads at finite bias voltage and the same temperature for both leads.
At $T=0$, an exact analytical expression is provided to third order
in $eV/(k_{B}T_{K})$. For finite temperature, a non monotonic behavior
of one of the currents is obtained as a function of $V$.

The paper is organized as follows. In Section \ref{model} we describe the
system and the IAM used to represent it. In Section \ref{rpt} we review
briefly the formalism of the RPT and obtain the analytical expressions for $%
\Sigma ^{<}$ and $\Sigma ^{>}$ for small energies. In Section \ref{nga} we
describe the Ng approximation. Section \ref{cc} contains a discussion on the
conservation of the current. In Section \ref{num} the results for $\Sigma
^{<}(\omega )$ calculated with RPT to second order in the renormalized
Coulomb repulsion are compared with the Ng approximation and the analytical
expression at different voltages and temperatures. 
In Section \ref{thermal} we show how the bias voltage originate heat currents.
Section \ref{disc} contains a summary and discussion.

\section{Model}

\label{model}

We use the IAM, to describe a semiconductor QD or a single molecule attached
to two conducting leads, with a bias voltage $V$ applied between these
leads. The Hamiltonian can be split into a noninteracting part $H_{0}$ and a
perturbation $H^{\prime }$ as \cite{none,lady} 
\begin{eqnarray}
H &=&H_{0}+H^{\prime },  \notag \\
H_{0} &=&\sum_{k\nu \sigma }\varepsilon _{k\nu }\,c_{k\nu \sigma }^{\dagger
}c_{k\nu \sigma }+\sum_{\sigma }\varepsilon _{\text{eff}}^{\sigma
}\,n_{d\sigma }  \notag \\
&&+\sum_{k\nu \sigma }\left( V_{k\nu }\,c_{k\nu \sigma }^{\dagger }d_{\sigma
}+\text{H.c.}\right) ,  \notag \\
H^{\prime } &=&\sum_{\sigma }\left( E_{d}-\sigma \mu _{B}B-\varepsilon _{%
\text{eff}}^{\sigma }\right) \,n_{d\sigma }+U\,n_{d\uparrow }n_{d\downarrow
},  \label{h}
\end{eqnarray}%
where $n_{d\sigma }=d_{\sigma }^{\dagger }d_{\sigma }$, and $\nu =L,R$
refers to the left and right leads. In general $\varepsilon _{\text{eff}%
}^{\sigma }$ is determined selfconsistently, except for the electron-hole
symmetric (EHS) case ($E_{d}=\mu -U/2$) with magnetic field $B=0$, for which 
$\varepsilon _{\text{eff}}^{\sigma }=\mu $,\cite{none,levy} where $\mu $ is
the Fermi level which we set as zero in the following.

We write the chemical potentials of both leads in the form

\begin{equation}
\mu _{L}=\alpha _{L}eV\text{, }\mu _{R}=-\alpha _{R}eV,  \label{mu}
\end{equation}%
where $\alpha _{L}+\alpha _{R}=1$. Similarly, the couplings to the leads
assumed independent of frequency are expressed in terms of the total
resonant level width $\Delta =\Delta _{L}+\Delta _{R}$ as (we take $\hbar =1$
in what follows)

\begin{equation}
\Delta _{\nu }=\pi \sum_{k}|V_{k\nu }|^{2}\delta (\omega -\varepsilon _{k\nu
})=\beta _{\nu }\Delta .  \label{del}
\end{equation}

\section{Renormalized perturbation theory}

\label{rpt}

The basic idea of RPT is to reorganize the perturbation expansion in terms
of fully dressed quasiparticles in a Fermi liquid picture.\cite{he1} The
parameters of the original model are renormalized and their values can be
calculated exactly from Bethe ansatz results, or accurately using NRG. One
of the main advantages is that the renormalized expansion parameter $%
\widetilde{U}/(\pi \widetilde{\Delta })$ is small. In the EHS case $%
\widetilde{U}/(\pi \widetilde{\Delta })\leq 1$, being 1 in the extreme Kondo
regime ($U=-2E_{d}\rightarrow \infty $).\cite{he1,ogu2} Within RPT, the low
frequency part of $G_{d\sigma }^{r}(\omega )$ is approximated as \cite{he1}

\begin{equation}
G_{d\sigma }^{r}(\omega )\simeq \frac{z}{\omega -\widetilde{\varepsilon }_{%
\text{eff}}^{\sigma }+i\widetilde{\Delta }-\widetilde{\Sigma }_{\sigma
}^{r}(\omega )},  \label{gra}
\end{equation}%
where $\widetilde{\Delta }=z\Delta $ is the renormalized resonant level
width, $z$ is the quasiparticle weight, $\widetilde{\varepsilon }_{\text{eff}%
}^{\sigma }$ is the renormalized level energy and $\widetilde{\Sigma }%
_{\sigma }^{r}(\omega )$ is the renormalized retarded self-energy (with $%
\widetilde{\Sigma }_{\sigma }^{r}(0)=\partial \widetilde{\Sigma }_{\sigma
}^{r}(\omega )/\partial \omega =0$ at $V=\omega =0$). $\widetilde{\Delta }$
is of the order of $k_{B}T_{K}$, where $T_{K}$ is the Kondo temperature.

The spectral density of $d$ electrons is $\rho _{\sigma }(\omega )=-\text{Im}%
G_{d\sigma }^{r}(\omega )/\pi $. The free quasiparticle spectral density of $%
d$ electrons is given by 
\begin{equation}
\widetilde{\rho }_{0}^{\sigma }(\omega )=\frac{\widetilde{\Delta }/\pi }{%
(\omega -\widetilde{\varepsilon }_{\text{eff}}^{\sigma })^{2}+\widetilde{%
\Delta }^{2}}.  \label{rhor}
\end{equation}%
Both densities at the Fermi energy can be related to the occupancy by the
Friedel sum rule \cite{lady,lan}

\begin{equation}
\pi \Delta \rho _{\sigma }(0)= \pi \widetilde{\Delta } \widetilde{\rho }%
^{\sigma}_{0}(0) =\sin ^{2}(\pi \langle n_{d\sigma} \rangle ),  \label{rho0}
\end{equation}
which allows one to relate the effective dot level with its occupancy

\begin{equation}
\widetilde{\varepsilon }_\text{eff}^{\sigma }=\widetilde{\Delta }\cot (\pi
\langle n_{d\sigma }\rangle ).  \label{eff}
\end{equation}

The lesser Green's function can be written in the form \cite{none,scali}

\begin{equation}
G_{d\sigma }^{<}(\omega )=\frac{|G_{d\sigma }^{r}(\omega )|^{2}}{z}\left( 2i%
\widetilde{\Delta }\tilde{f}(\omega )-\widetilde{\Sigma }_{\sigma
}^{<}(\omega )\right) ,  \label{gl}
\end{equation}%
where 
\begin{equation}
\tilde{f}(\omega )=\sum_{\nu }\beta _{\nu }f(\omega -\mu _{\nu })
\label{fav}
\end{equation}%
is a weighted average of the Fermi functions $f(\omega )=1/(e^{\omega
/k_{B}T}+1)$ at the two leads, and $\widetilde{\Sigma }_{\sigma }^{<}(\omega
)$ is the renormalized lesser self-energy.

The greater quantities can be obtained from the retarded and lesser ones
using the relations \cite{lif}

\begin{eqnarray}
G^<-G^>=G^a-G^r=-2i \text{Im}G^r(\omega),  \label{gdif} \\
\Sigma^<-\Sigma^>=\Sigma^a-\Sigma^r=2i \text{Im}\Sigma^r(\omega),
\label{sdif}
\end{eqnarray}
where we have used that in the frequency domain, the advanced quantities $%
G^a(\omega)$, $\Sigma^a(\omega)$ are the complex conjugates of the
corresponding retarded ones.

In the following, we assume that $B=0$ and the leads are paramagnetic, so
that the subscript $\sigma $ can be dropped, and $\langle n_{d\sigma
}\rangle =n/2$, where $n$ is the total occupancy at the QD.

The linear term in the specific heat and the impurity contribution to the
magnetic susceptibility at zero temperature are given by \cite{he1}

\begin{align}
\gamma _{C}& =2\pi ^{2}k_{B}^{2}\widetilde{\rho }_{0}(0)/3,  \label{gam} \\
\chi & =(g\mu _{B})^{2}\widetilde{\rho }_{0}(0)(1+\widetilde{U}\widetilde{%
\rho }_{0}(0))/2,  \label{xi}
\end{align}

These equations can be inverted to obtain the effective parameters from 
an accurate knowledge of thermodynamic quantities. For example from Eqs. (%
\ref{rho0}), (\ref{eff}) and (\ref{gam})

\begin{equation}
\widetilde{\Delta }=\frac{2\pi k_{B}^{2}}{3\gamma _{C}}\sin ^{2}(\pi n/2),
\label{delt}
\end{equation}
and the renormalized interaction is obtained through the Wilson ratio

\begin{equation}
R=\frac{\chi }{\gamma _{C}}\frac{1}{3}\left( \frac{2\pi k_{B}}{g\mu _{B}}%
\right) ^{2}=1+\widetilde{U}\widetilde{\rho }_{0}(0)  \label{wr}
\end{equation}

\subsection{Renormalized lesser and greater self-energies}

\label{sigl}

The renormalized self-energies are calculated as in ordinary perturbation
theory in the Keldysh formalism using the low-energy approximation for the
unperturbed Green functions.\cite{he1,ogu1,ogu2} To order $\widetilde{U}^{2}$%
, the renormalized lesser and greater self-energies can be written as \cite%
{none}

\begin{eqnarray}
\widetilde{\Sigma}^{<}(\omega ) &=&z\Sigma ^{<}(\omega )  \notag \\
&=&-2 \pi i \widetilde{U}^{2}\int d\epsilon _{1}d\epsilon _{2} \widetilde{%
\rho }_{0}(\epsilon _{1})\widetilde{\rho }_{0}(\epsilon _{2}) \widetilde{%
\rho }_{0}(\epsilon _{1}+\epsilon _{2}-\omega )  \notag \\
&&\times \tilde{f}(\epsilon _{1})\tilde{f}(\epsilon _{2}) (1-\tilde{f}%
(\epsilon _{1}+\epsilon _{2}-\omega )),  \label{slro}
\end{eqnarray}

\begin{eqnarray}
\widetilde{\Sigma}^{>}(\omega ) &=& 2 \pi i \widetilde{U}^{2}\int d\epsilon
_{1}d\epsilon _{2}\widetilde{\rho }_{0}(\epsilon _{1})\widetilde{\rho }_{0}
(\epsilon _{2})\widetilde{\rho }_{0}(\epsilon _{1}+\epsilon_{2}-\omega ) 
\notag \\
&&\times (1-\tilde{f}(\epsilon _{1}))(1-\tilde{f}(\epsilon _{2})) \tilde{f}%
(\epsilon _{1}+\epsilon _{2}-\omega ).  \label{sgro}
\end{eqnarray}

\subsubsection{Analytical approximation for small energies}

\label{ana} In this Section we calculate the lesser and greater
self-energies assuming that the energies $\omega $, $eV$ and $k_{B}T$ are
small in comparison with $\widetilde{\Delta }$, which in turn is of the
order of $k_{B}T_{K}$.\cite{he1} Specifically, to evaluate the self-energies
to total second order in $\omega $, $V$ and $T$, it suffices to replace the
quasiparticle spectral density $\widetilde{\rho }_{0}(\epsilon )$ by its
value at the Fermi energy (order 0 in an expansion in $\omega $), because
the two integrations in Eqs. (\ref{slro}) and (\ref{sgro}) already introduce
terms of second order, due to the effect of the Fermi functions in
restricting the intervals of $\epsilon _{i}$ for which the integrand has non
negligible values. For the same reason, terms of higher order in $\widetilde{%
U}$ lead to terms of higher order in $\omega $, $V$ or $T$. Therefore, the
result below is exact to second order. Note that besides the evaluation to
second order in $\widetilde{U}$, the only additional approximation is
neglecting the energy dependence of $\widetilde{\rho }_{0}(\epsilon )$. The
Fermi functions are treated exactly and are not expanded.\cite{note2}

Using Eq. (\ref{fav}) one sees that 
\begin{equation}
\tilde{f}(x)+\tilde{f}(-x)=f(x)+f(-x)=1,  \label{one}
\end{equation}
which together with Eq. (\ref{fav}) allows to write the approximation of Eq.
(\ref{slro}) for small arguments as

\begin{eqnarray}
\widetilde{\Sigma}_2^{<}(\omega,V,T) &=&-2ip\sum_{\nu \xi \kappa }\beta
_{\nu }\beta _{\xi }\beta _{\kappa }\int d\epsilon _{1}d\epsilon
_{2}f(\epsilon _{1}-\mu _{\nu })  \notag \\
&&\times f(\epsilon _{2}-\mu _{\xi }) f(\omega +\mu _{\kappa }-\epsilon
_{1}-\epsilon _{2}),  \label{sl2}
\end{eqnarray}
where the factor

\begin{eqnarray}
p &=&\pi \lbrack \widetilde{\rho }_{0}(0)]^{3}\widetilde{U}^{2}=\frac{%
(R-1)^{2}\sin ^{2}(\pi n/2)}{\widetilde{\Delta }}  \notag \\
&=&\frac{3(R-1)^{2}\gamma _{C}}{2\pi k_{B}^{2}}=\frac{2\pi (R-1)^{2}\chi }{%
R(g\mu _{B})^{2}},  \label{p}
\end{eqnarray}%
can be expressed in terms of the linear term in the specific heat and the
magnetic susceptibility at $T=0$.

The integrals in Eq. (\ref{sl2}) are evaluated analytically as described in
the appendix. The result can be written in the form

\begin{eqnarray}
\widetilde{\Sigma }_{2}^{<} &=&-ip\sum_{j}c_{j}f(a_{j})\left[ a_{j}^{2}+(\pi
k_{B}T)^{2}\right] ,  \notag \\
c_{1} &=&\beta _{L}^{2}\beta _{R},\text{ }a_{1}=\omega -(1+\alpha _{L})eV, 
\notag \\
c_{2} &=&\beta _{L}^{3}+2\beta _{L}\beta _{R}^{2},\text{ }a_{2}=\omega
-\alpha _{L}eV,  \notag \\
c_{3} &=&\beta _{R}^{3}+2\beta _{L}^{2}\beta _{R},\text{ }a_{3}=\omega
+\alpha _{R}eV,  \notag \\
c_{4} &=&\beta _{L}\beta _{R}^{2},\text{ }a_{4}=\omega +(1+\alpha _{R})eV.
\label{sl3}
\end{eqnarray}%
Particular cases of this low-energy expansion were derived before.\cite%
{hersh,scali} Using Eqs. (\ref{sgro}), (\ref{one}), (\ref{sl2}) and (\ref%
{fav}) one sees that to total second order in $\omega $, $V$ and $T$, the
greater self-energy becomes simply

\begin{equation}
\widetilde{\Sigma}_2^{>}(\omega, V, T)=-\tilde{\Sigma}_2^{<}(-\omega, -V, T).
\label{sg2}
\end{equation}

It is interesting to note that to the same order, calculating the imaginary
part of $\widetilde{\Sigma}^{r}$ from the difference Eq. (\ref{sdif}), using
Eqs. (\ref{one}), (\ref{sl3}) and (\ref{sg2}), the Fermi functions disappear
and collecting the different terms one recovers the very simple result \cite%
{ogu2}

\begin{eqnarray}
\text{Im}\widetilde{\Sigma}_2 ^{r} &=&-\frac{p}{2}[\omega ^{2}-2\gamma
\omega eV+\delta (eV)^{2}+(\pi k_{B}T)^{2}],  \label{sret} \\
\gamma &=&\alpha _{L}\beta _{L}-\alpha _{R}\beta _{R},  \label{gamma} \\
\delta &=&\gamma ^{2}+3\beta _{L}\beta _{R}.  \label{delta}
\end{eqnarray}

\subsubsection{Ward identities}

\label{ward}

The different self-energies should satisfy the Ward identities \cite%
{ogu1,ogu2} 
\begin{equation}
{\frac{\partial \tilde{\Sigma}^{\eta }(\omega )}{\partial eV}\vline}%
_{V=0}=-\gamma \left( \frac{\partial \tilde{\Sigma}^{\eta }(\omega )}{%
\partial \omega }+\frac{\partial \tilde{\Sigma}^{\eta }(\omega )}{\partial
E_{d}}\right) ,  \label{ds}
\end{equation}%
where the superscript $\eta $ denotes $>$, $<$, $r$ or $a$, and $\gamma $ is
given by Eq. (\ref{gamma}). These identities come simply from the properties
of the Fermi functions $f(\omega -\mu _{\nu })$ and evaluation at $V=0$
renders both of them equal after derivation [see Eqs. (\ref{mu})]. They are
satisfied at any order in perturbation theory.

Direct differentiation of the analytical expression (\ref{sl3}) gives 
\begin{eqnarray}
{\frac{\partial i\tilde{\Sigma}_{2}^{<}(\omega )}{\partial \omega }\vline}%
_{V=0} &=&p\frac{\omega }{1+e^{x}}\left( 2-\frac{x}{1+e^{-x}}\right) , 
\notag \\
x &=&\frac{\omega }{k_{B}T}  \label{dsw} \\
{\frac{\partial \tilde{\Sigma}_{2}^{<}(\omega )}{\partial eV}\vline}_{V=0}
&=&-\gamma \frac{\partial \tilde{\Sigma}_{2}^{<}(\omega )}{\partial \omega }.
\label{dg}
\end{eqnarray}%
$\partial \tilde{\Sigma}_{2}^{<}/\partial E_{d}$ can be neglected since it
only modifies $\widetilde{\rho }_{0}(0)$ and therefore leads to a
contribution of higher order. Thus, $\tilde{\Sigma}_{2}^{<}$ satisfies the
Ward identity Eq. (\ref{ds}) to linear order in $\omega $ and $\omega x$.
These results will be discussed further in Section \ref{cr}. The $%
T\rightarrow 0$ limit is well defined and the Ward identity is also
satisfied by $\tilde{\Sigma}_{2}^{<}(\omega )$ at $T=0$ in spite of the
claim in Ref. \onlinecite{reply} that it is not the case.\cite{rr,note2}

It is trivial to see that $\text{Im}\tilde{\Sigma}_2^{r}$ [Eq. (\ref{sret})]
also satisfies Eq. (\ref{ds}) and from Eq. (\ref{sdif}), $\tilde{\Sigma}%
_2^{>}$ satisfies the Ward identity too.

\section{Ng approximation}

\label{nga}

The Ng approximation can be written as 
\begin{equation}
\widetilde{\Sigma }_{\text{Ng}}^{<}(\omega )=2i\tilde{f}(\omega ) \text{Im}%
\widetilde{\Sigma }^{r}(\omega ),  \label{ngsl}
\end{equation}%
where $\tilde{f}(\omega )$ is defined by Eq. (\ref{fav}). Using Eq. (\ref{gl}%
) it can be written in the equivalent form 
\begin{equation}
G_{\text{Ng}}^{<}(\omega )=-2i\tilde{f}(\omega )\text{Im}G^{r}(\omega ),.
\label{nggl}
\end{equation}%
Using Eqs. (\ref{gdif}) and (\ref{sdif}) also the greater quantities become
proportional to the retarded ones:

\begin{eqnarray}
\widetilde{\Sigma }_{\text{Ng}}^{>}(\omega ) &=&-2i[1-\tilde{f}(\omega )]%
\text{Im}\widetilde{\Sigma }^{r}(\omega ),  \label{ngsg} \\
G_{\text{Ng}}^{>}(\omega ) &=&2i[1-\tilde{f}(\omega )]\text{Im}G^{r}(\omega
),.  \label{nggg}
\end{eqnarray}

These equations are exact in the non-interacting case ($U=0$) and also at
equilibrium ($V=0$).\cite{ng} In addition using the results of RPT\ up to $%
\widetilde{U}^{2}$ for $\widetilde{\Sigma }^{r}(\omega )$, it can be shown
that at $T=0$, the perturbative result $\widetilde{\Sigma }^{<}(\omega )$
and the corresponding Ng approximation $\widetilde{\Sigma }_{\text{Ng}%
}^{<}(\omega )$ coincide for $\omega <$ -$(1+\alpha _{R})eV$ and $\omega
>(1+\alpha _{L})eV$. However, if the expression Eq. (\ref{sret}) for $\text{%
Im}\widetilde{\Sigma }^{r}(\omega )$ at small energies is replaced in Eq. (%
\ref{ngsl}), an analytical expression for $\widetilde{\Sigma }_{\text{Ng}%
}^{<}(\omega )$ is obtained which is obviously different from the exact
result for small $\omega $, $V$ and $T$, Eq. (\ref{sl3}). The quantitative
differences will be discussed in Section \ref{num}.

\section{Conservation of the current}

\label{cc}

Using the Keldysh formalism,\cite{past,meir} the current flowing between the
left lead and the dot can be written as

\begin{equation}
I_{L}=\frac{4ie\Delta _{L}}{h}\int d\omega \left[ 2if(\omega -\mu _{L}) 
\text{Im}G_{d}^{r}(\omega )+G_{d}^{<}(\omega )\right],  \label{il}
\end{equation}%
while the current flowing between the dot and the right lead is

\begin{equation}
I_{R}=-\frac{4ie\Delta _{R}}{h}\int d\omega \left[ 2if(\omega -\mu _{R})%
\text{Im}G_{d}^{r}(\omega )+G_{d}^{<}(\omega )\right].  \label{ir}
\end{equation}
Conservation of the current requires $I_{L}=I_{R}=I$.

Using Eqs. (\ref{gra}) and (\ref{gl}), the difference can be written in the
form

\begin{equation}
I_{L}-I_{R}=-\frac{4e\widetilde{\Delta }}{h}\int d\omega {\vline\frac{%
G_{d}^{r}(\omega )}{z}\vline}^{2}[2\tilde{f}(\omega )\text{Im}\widetilde{%
\Sigma }^{r}(\omega )+i\tilde{\Sigma}^{<}(\omega )].  \label{dif}
\end{equation}%
Using Eqs. (\ref{sl3}) and (\ref{sret}), it is easy to see that to total
third order in $eV/\widetilde{\Delta }$ and $k_{B}T/\widetilde{\Delta }$
this expression vanishes. Thus RPT conserves the current to this order.
Instead, if Ng approximation Eq. (\ref{ngsl}) is used, $I_{L}-I_{R}$
vanishes identically and the current is conserved to all orders.

\section{Lesser self-energy to second order in $\widetilde{U}$}

\label{num}

In this Section we present results for $\tilde{\Sigma}^{<}(\omega )$
calculated with RPT to second order in $\widetilde{U}$ by numerical
integration. The expression used is equivalent to Eq. (\ref{slro}) but we
have used a different approach explained in the appendix of Ref. %
\onlinecite{none}, in which one integral is evaluated analytically. This
result $\tilde{\Sigma}^{<}(\omega )$ is superior to the analytical one $%
\tilde{\Sigma}_{2}^{<}(\omega )$ [Eq. (\ref{sl3})] because no additional
approximations (constant quasiparticle density) were made. Both coincide to
total second order in $\omega $, $V$ and $T$. Therefore the difference is
due to higher order terms in $\tilde{\Sigma}^{<}(\omega )$.

For the calculation of the current, we also need the real part of the
renormalized retarded self-energy $\tilde{\Sigma}^{r}(\omega )$, which is
also calculated as in Ref. \onlinecite{none} with the constant and linear
terms in $\omega $ for $V=T=0$ subtracted.\cite{he1,scali}

We have chosen a total occupation $n=2\langle n_{d\sigma }\rangle =3/4$ (out
of the EHS case). From Eq. (\ref{eff}) this implies $\widetilde{\varepsilon }%
_{\text{eff}}^{\sigma }=(\sqrt{2}-1)\widetilde{\Delta }$. We have taken $%
\widetilde{U}/(\pi \widetilde{\Delta })=1$ for simplicity.\cite{note} This
quotient enters as a constant factor $[\widetilde{U}/(\pi \widetilde{\Delta }%
)]^{2}$ in $\tilde{\Sigma}^{<}(\omega )$ but modifies the values of the
current discussed below. Preliminary NRG results indicate that for 
$E_{d}=-2\Delta $ and $U\rightarrow +\infty $, one has $n=3/4$ and
renormalized parameters $z=\widetilde{\Delta }/\Delta =0.115$ and 
$\widetilde{U}/(\pi \widetilde{\Delta })=1.136$.\cite{aap}

We assume here a symmetric voltage drop $\alpha _{L}=\alpha _{R}=1/2.$This
is motivated by the fact that even for molecular quantum dots with high
asymmetric coupling to the leads ($\beta _{L}\gg \beta _{R}$ or $\beta
_{L}\ll \beta _{R}$), the shape of the diamonds with the regions of high
conductivity as a function of bias voltage $V$ and gate voltage $V_{g}$
indicates a rather symmetric voltage drop. Instead, we consider different
ratios of $\beta _{L}/\beta _{R}$.

\subsection{Symmetric coupling to the leads}

\label{cs}

\begin{figure}[h]
\includegraphics[width=8.cm]{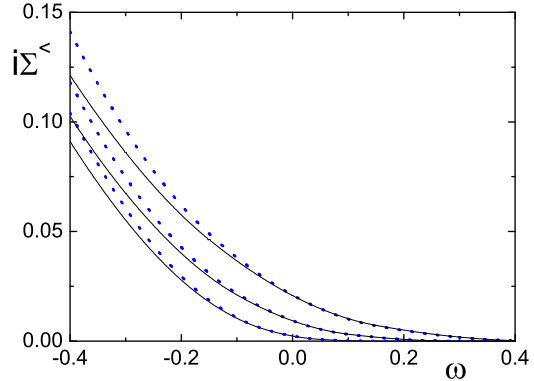}
\caption{(Color online) Full lines: renormalized lesser self-energy as a
function of frequency for $\protect\beta_L=\protect\beta_R$, $T=0$ and
several bias voltages. From bottom to top $eV=$ 0.1, 0.2 and 0.3. Dotted
line: analytical result at small energies [Eq. (\protect\ref{sl3})]. $%
\widetilde{\Delta }=1$ is taken as the unit of energy.}
\label{sv}
\end{figure}

In Fig. \ref{sv} we show $\widetilde{\Sigma }^{<}(\omega )$ for $\beta
_{L}=\beta _{R}$ and different values of $V$ at zero temperature. In the
equilibrium case $V=0$ (not shown), it is known that $\widetilde{\Sigma }%
^{<}(\omega )=2if(\omega ) \text{Im} \widetilde{\Sigma }^{r}(\omega )$, $%
\widetilde{\Sigma }^{r}(\omega )\sim \omega ^{2}$ for small $\omega $ [Eq. (%
\ref{sret})] and therefore, $i\widetilde{\Sigma }^{<}(\omega )$ is a
decreasing function of $\omega $ for negative $\omega $ and zero for
positive $\omega $ at $T=0$. The expression Eq. (\ref{sl3}) indicates that
the effect of a small voltage is to split this result into four similar
expressions, two shifted to smaller $\omega $ and two to higher $\omega .$
The net effect is to increase $i\widetilde{\Sigma }^{<}(\omega )$, but it
continues to be a monotonically decreasing function.

The comparison between the numerical result $\widetilde{\Sigma }^{<}(\omega) 
$ and the analytical one $\widetilde{\Sigma }_{2}^{<}(\omega )$ [Eq. (\ref%
{sl3})] to total second order in $\omega $ and $V$ is good for $|\omega |<0.2%
\widetilde{\Delta }$, suggesting that higher order terms are small in this
interval. Instead, for $-\omega <0.2\widetilde{\Delta }$, $\widetilde{\Sigma 
}_{2}^{<}(\omega )$ overestimates $\widetilde{\Sigma }^{<}(\omega )$.

We have also calculated the currents between the left lead and the dot $%
I_{L} $ and between the dot and the right lead $I_{R}$ for $eV\leq 0.4 
\widetilde{\Delta }$. The relative error $|I_{L}-I_{R}|/I$, where $%
I=(I_{L}+I_{R})/2,$ is less than $2.2 \times 10^{-4}$ for the values of $eV$
studied. An excellent fit of the difference between currents in this
interval is $I_{L}-I_{R}=(2e/h)[-0.00311(eV/\widetilde{\Delta }%
)^{4}-0.00777(eV/\widetilde{\Delta })^{5}]$.\cite{note} This confirms the
analysis of the previous section that the current is conserved to order $%
V^{3}$ by RPT. In the same interval the current can be fitted by $%
I=(2e/h)[0.8531(eV/\widetilde{\Delta })-0.1754(eV/\widetilde{\Delta })^{3}]$%
. The linear term agrees with the expected conductance from Friedel sum
rule, proportional to $\sin ^{2}(\pi \langle n_{d\sigma }\rangle )=(2+\sqrt{2%
})/4\approx 0.8536$.

\begin{figure}[h]
\includegraphics[width=8.cm]{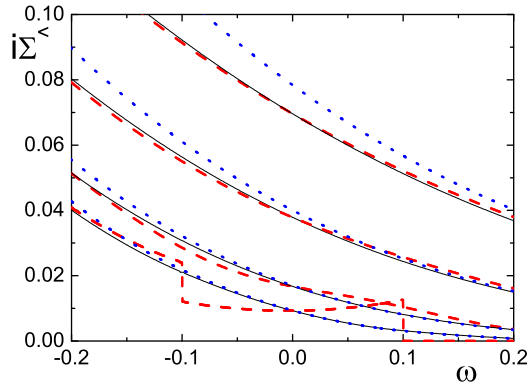}
\caption{(Color online) Full lines: renormalized lesser self-energy as a
function of frequency for $\protect\beta_L=\protect\beta_R$, $eV=0.2$ and
several temperatures. From bottom to top $k_BT=$ 0, 0.05, 0.1 and 0.2.
Dashed line: Ng approximation [Eq. (\protect\ref{ngsl})]. Dotted line:
analytical result at small energies [Eq. (\protect\ref{sl3})].}
\label{st}
\end{figure}

The effect of temperature on $i\widetilde{\Sigma }^{<}(\omega )$ is shown
Fig. \ref{st} and the result is compared with the analytical expression for
small $\omega $, $V$ and $T$ [Eq. (\ref{sl3})] and the Ng approximation [Eq.
(\ref{ngsl})]. While as shown above, the former expression $\widetilde{%
\Sigma }_{2}^{<}(\omega )$ works well at $T=0$, the Ng approximation $%
\widetilde{\Sigma }_{\text{Ng}}^{<}(\omega )$ fails in the region of small
frequencies, below $eV$. In particular, it has jumps at both chemical
potentials $\mu _{\nu }$ due to the factor $\tilde{f}(\omega )$ [Eq. (\ref%
{fav})] in Eq. (\ref{ngsl}) and it increases in some interval at positive
frequencies in contrast to the overall decreasing behavior of $i\widetilde{%
\Sigma }^{<}(\omega )$. However, the Ng approximation improves rapidly with
increasing temperature. For $k_{B}T=eV/4$, $i\widetilde{\Sigma }_{\text{Ng}%
}^{<}(\omega )$ lies a little bit below (above) $i\widetilde{\Sigma }%
^{<}(\omega )$ for $\omega $ near to the smaller (greater) chemical
potential. For $k_{B}T=eV/2$, $\widetilde{\Sigma }_{\text{Ng}}^{<}(\omega )$
is already a good approximation for $\widetilde{\Sigma }^{<}(\omega )$ in
the whole frequency range. Instead, the analytical expression $\widetilde{%
\Sigma }_{2}^{<}(\omega )$ overestimates $\widetilde{\Sigma }^{<}(\omega )$
for $k_{B}T\geq eV/2$, particularly at negative frequencies, indicating that
terms in temperature of higher order than $T^{2}$ become important.

Concerning the conservation of the current, $|I_L-I_R|/I$ remains below $%
0.001$ for $eV=0.2 \widetilde{\Delta }$ and $k_{B}T \leq \widetilde{\Delta }$%
.

\subsection{Larger coupling to the lead of higher chemical potential}

\label{cl}

\begin{figure}[h!]
\includegraphics[width=8.cm]{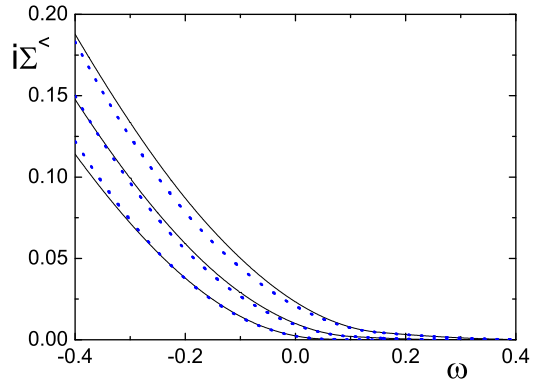}
\caption{(Color online) Same as Fig. \protect\ref{sv} for $\protect\beta_L=9%
\protect\beta_R$.}
\label{slv}
\end{figure}

In this Section we study the case $\beta _{L}=9\beta _{R}$. As seen in Fig. %
\ref{slv}, increasing the coupling with the left lead, for which the
chemical potential $\mu _{L}=\alpha _{L}eV>0$ has the main effect of
shifting $i\widetilde{\Sigma }^{<}(\omega )$ to higher frequencies. Since $i%
\widetilde{\Sigma }^{<}(\omega )$ is a decreasing function of $\omega $,
this shift implies higher values $i\widetilde{\Sigma }^{<}(\omega )$ for
fixed $\omega $. This can be understood from the analytical expression Eq. (%
\ref{sl3}) in which the terms with coefficients $c_{1}$ and $c_{2}$ increase
in magnitude. For $\beta _{L}\rightarrow 1$ ($\beta _{R}\rightarrow 0$),
only $c_{2}$ survives and all self-energies reduce to those of a QD at
equilibrium with the left lead, $\text{Im}\widetilde{\Sigma }^{r}(\omega )$
behaves as $(\omega -\mu _{L})^{2}$ for small $\omega $ and $V$ at $T=0$
[see Eq. (\ref{sret})], the Ng approximation becomes exact and $\tilde{f}%
(\omega )=f(\omega -\mu _{L})$. While this limit is still not reached for $%
\beta _{L}=9\beta _{R}$, one expects a smaller ratio $|I_{L}-I_{R}|/I$ and a
better comparison with the Ng approximation. However, while the currents
decrease, the ratio $|I_{L}-I_{R}|/I$ is of the same order of magnitude as
before, for the range of voltages studied. The same happens for the case $%
\beta _{L}=\beta _{R}/9$ discussed in Section \ref{cr}.

\begin{figure}[h!]
\includegraphics[width=8.cm]{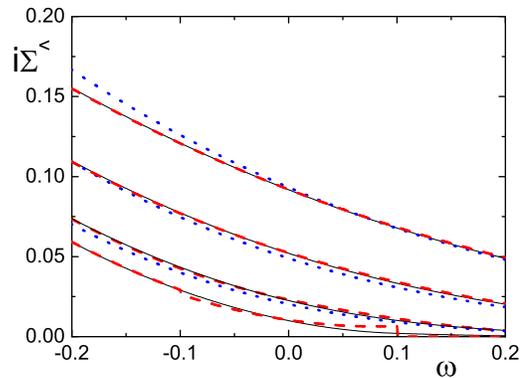}
\caption{(Color online) Same as Fig. \protect\ref{st} for $\protect\beta_L=9%
\protect\beta_R$.}
\label{slt}
\end{figure}

The evolution of $\widetilde{\Sigma }^{<}(\omega )$ with temperature is
shown in Fig. \ref{slt} and compared with Ng and analytical approximations.
At zero temperature, $\widetilde{\Sigma }_{\text{Ng}}^{<}(\omega )$ has
qualitatively similar shortcomings as for symmetric coupling to the leads,
with jumps at both $\mu _{\nu }$, but quantitatively the agreement is
better, as expected. At finite temperature, in this case, already for $%
k_{B}T=eV/4$, the Ng approximation reproduces very well $\widetilde{\Sigma }%
^{<}(\omega )$. For higher temperatures the agreement improves, while the
analytical approximation $\widetilde{\Sigma }_{2}^{<}$ becomes worse.

\subsection{Larger coupling to the lead of lower chemical potential}

\label{cr}

\begin{figure}[h!]
\caption{(Color online) Renormalized lesser self-energy as a function of
frequency for $\protect\beta_L=\protect\beta_R/9$, $T=0$ and several bias
voltages indicated inside the figure. }
\label{srv}\includegraphics[width=8.cm]{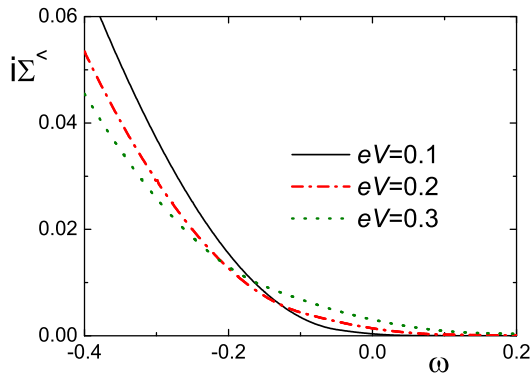}
\end{figure}

In this Section we consider the opposite case as in Section \ref{cl} and
take $\beta _{L}=\beta _{R}/9$. In this case, the system is nearer to the
situation in which the dot is at equilibrium with the right lead and similar
considerations as in the previous Section apply. In Fig. \ref{srv} we
display $\widetilde{\Sigma }^{<}(\omega )$ for several values of $V$. While
for small $\omega $, $i\widetilde{\Sigma }^{<}(\omega )$ increases with $V$,
the behavior changes for $-\omega >eV$ and $i\widetilde{\Sigma }^{<}(\omega
) $ decreases with increasing $V$. This can be understood from the Ward
identity [Eqs. (\ref{dg}) and (\ref{gamma}) for small $\omega $ and $V$].
While the identity is strictly valid for $V=0$ one expects it to be
qualitatively valid for small $eV$ compared to $|\omega |$. Since $\partial i%
\tilde{\Sigma}_{2}^{<}/\partial \omega |_{V=0}$ is negative for negative $%
\omega $ and also $\gamma $ is negative for large $\beta _{R}$, one expects
a decrease of $i\widetilde{\Sigma }^{<}(\omega )$ with increasing $V$ for $%
eV\ll -\omega $, as observed in Fig. \ref{srv}.

\begin{figure}[h!]
\includegraphics[width=8.cm]{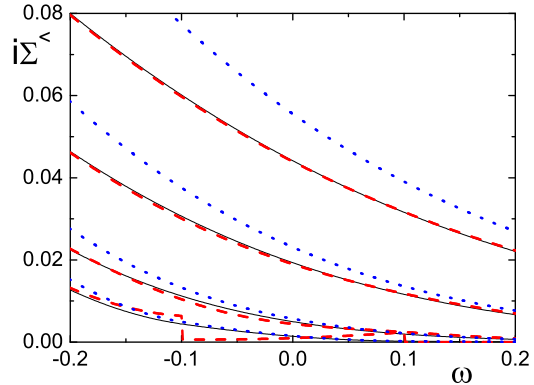}
\caption{(Color online) Same as Fig. \protect\ref{st} for $\protect\beta_L=%
\protect\beta_R/9$.}
\label{srt}
\end{figure}

The effect of temperature on $i\widetilde{\Sigma }^{<}(\omega )$ is shown in
Fig. \ref{srt}. The deviations at zero temperature between the Ng
approximation and the correct result to order $\widetilde{U}^{2}$ are larger
than in the previous case, particularly for $\omega $ near $\mu _{R}$ ($-0.1%
\widetilde{\Delta }$ in the figure). However, the comparison improves
rapidly with increasing temperature, and $\widetilde{\Sigma }_{\text{Ng}%
}^{<}(\omega )$ turns out to be a good approximation for $k_{B}T\geq eV/4.$

\section{Thermal current induced by the voltage}

\label{thermal}

In this Section, we discuss the heat currents $J_{L}$ flowing from the left
lead to the dot and  $J_{R}$ flowing from the dot to the right lead. From
the thermodynamic equation $dQ=dE-\mu N$, it is clear that 

\begin{equation}
J_{\nu }=J_{\nu }^{E}-\mu _{\nu }J_{\nu }^{N},  \label{jq}
\end{equation}%
where $J_{\nu }^{E}$ are the energy currents and $J_{\nu }^{N}$ are the
corresponding particle currents.

For a model with nearest-neighbor hopping only, an energy density can be
defined and using the continuity equation the energy current can be defined.\cite{loza} 
Alternatively, following the definition given by Boese and Fazio 
\cite{boese} and using the formalism of Meir and Wingreen,\cite{meir} one
arrives at the same expressions, similar to Eqs. (\ref{il}) and (\ref{ir})

\begin{equation}
J_{\nu }^{E}=\pm \frac{4i\Delta _{L}}{h}\int \omega d\omega \left[
2if(\omega -\mu _{\nu })\text{Im}G_{d}^{r}(\omega )+G_{d}^{<}(\omega )\right]
,  \label{je}
\end{equation}%
where upper (lower) sign corresponds to $\nu =L$ ($R$). 
These expressions were obtained previously by Dong and Lei,\cite{dong}
who calculated the thermopower of a quantum dot in the linear response regime 
($V \rightarrow 0$ and vanishing temperature gradient) using Ng ansatz for 
$G_{d}^{<}(\omega )$.

The energy current is conserved:  $J_{L}^{E}=J_{R}^{E}$. Following a similar
reasoning as in Section \ref{cc}, it is easily seen that this condition is
satisfied to total fourth order in in $eV/\widetilde{\Delta }$ and $k_{B}T/%
\widetilde{\Delta }$ by the RPT expressions and exactly by the Ng
approximation. Adding the first Eq. (\ref{je}) for times $\Delta _{L}$ plus the
second times $\Delta _{R}$ and using  $J_{L}^{E}=J_{R}^{E}$, an expression
for the energy current is obtained in which $G_{d}^{<}(\omega )$ is
eliminated. The same trick has been used for the electric 
currents,\cite{meir}  which are the particle currents times the elementary charge: 
$I_{\nu}=eJ_{\nu }^{N}$. Using this and  Eqs. (\ref{il}) and (\ref{ir}) one
obtains

\begin{equation}
J_{\nu }=\frac{8\pi \beta _{L}\beta _{R}\Delta }{h}\int (\omega -\mu _{\nu
})d\omega \rho (\omega )[f_{L}(\omega )-f_{R}(\omega )].  \label{jq2}
\end{equation}%
Note that the heat current is not conserved. The difference 
$J_{R}-J_{L}=(\mu _{L}-\mu _{R})I_{v}/e=I_{v}V$ is precisely the Joule
heating at the quantum dot. 

At zero temperature, the exact heat currents to order $(eV/\widetilde{\Delta 
})^{3}$ can be obtained using Eq. (\ref{rho0}) and \cite{scali}

\begin{equation}
\frac{\rho (\omega )}{\rho (0)}\simeq 1+\sin (\pi n)\left[ \frac{\omega
-\gamma (R-1)eV}{\widetilde{\Delta }}\right] .  \label{rhoap}
\end{equation}%
The result is

\begin{eqnarray}
&&J_{\nu } \simeq \frac{8\beta _{L}\beta _{R}}{h}(eV)^{2}\sin ^{2}(\pi n/2) 
\notag \\
&\times& \{\frac{\alpha _{L}-\alpha _{R}}{2}+\frac{eV\sin (\pi n)}
{\widetilde{\Delta }}[\frac{\alpha _{L}^{3}+\alpha _{R}^{3}}{3} \notag \\
&&-\frac{\gamma (R-1)(\alpha _{L}-\alpha _{R})}{2}]  \notag \\
&\mp & \alpha _{\nu }(1+\frac{eV\sin (\pi n)}{\widetilde{\Delta }}[\frac{%
\alpha _{L}-\alpha _{R}}{2}-\gamma (R-1)])\}.  \label{jq3}
\end{eqnarray}
The leading term gives $J_R=-J_L=G(0)V^2/2$, where $G(0)=8 \beta _{L}\beta _{R} \sin ^{2}(\pi n/2) e^2/h$ 
is the conductance at $V=0$.\cite{scali} Thus, for small $V$ the heat flow to each lead 
is the same independently of the particular voltage drops and coupling to the leads. 

\begin{figure}[h!]
\includegraphics[width=8.cm]{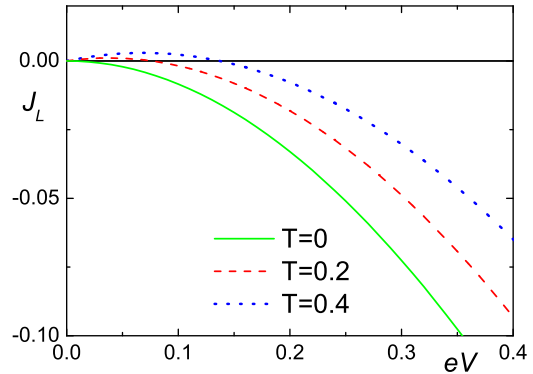}
\caption{(Color online) Thermal current between the left lead and the quantum dot
in units of $\widetilde{\Delta }^2/h$ for several temperatures. $\widetilde{\Delta }=1$
is the unit of energy. Parameters in the text.}
\label{jql}
\end{figure}

An analysis of the heat current in the general non-equilibrium case, with
different temperatures of the two leads, would require to perform
numerically three integrations in frequency. This is highly demanding.
Here we study the effect of temperature on the heat current assuming that it is the same 
for both leads.  We have taken $\widetilde{U}/(\pi \widetilde{\Delta })=1.136$. 
This value was obtained from recent NRG calculations for  
$E_{d}=-2\Delta $ and $U\rightarrow +\infty $, which also lead to $n=0.75$ 
and $z=\widetilde{\Delta }/\Delta =0.115$.\cite{aap} The result for $J_L$ 
for symmetric coupling to the leads and voltage drops $\alpha_{\nu}=\beta_{\nu}=1/2$ is shown in Fig. 
\ref{jql}. While for $T=0$, $J_L$ is negative, as expected from the leading 
quadratic term in Eq. (\ref{jq3}), the temperature leads to a positive linear term in $V$
(for both heat currents $J_{\nu}$) which dominates the current for small $V$. 
This positive contribution is expected in linear response, and is consistent
with the negative Seebeck coefficient $S$ for temperatures below the Kondo temperature reported 
previously at equilibrium for $n<1$ ($S$ is proportional to minus the energy current).\cite{vel,see}
As a consequence, for finite temperatures, $J_L$ changes sign as a function of the applied bias 
voltage. For occupation $n>1$, $S$ is positive and $J_R$ changes sign from negative to positive 
with increasing bias voltage. 

\section{Summary and discussion}

\label{disc}

Using renormalized perturbation theory (RPT) to second order in the
renormalized Coulomb repulsion $\widetilde{U}$, we have calculated the
lesser self-energy $\widetilde{\Sigma }^{<}(\omega )/z$ for the impurity
Anderson model, which describes transport through quantum dots, in the
general case (without electron-hole symmetry, asymmetric voltage drops and
different coupling to the conducting leads). The greater self-energy can be
calculated from the difference with the imaginary part of the retarded
self-energy [Eq. (\ref{sdif})]. Using an additional approximation, valid for
small $\hbar \omega /\widetilde{\Delta }$, $eV/\widetilde{\Delta }$ and $%
k_{B}T/\widetilde{\Delta }$, where $\widetilde{\Delta }/k_{B}$ is of the
order of the Kondo temperature $T_{K}$, we have derived exact analytical
expressions to to total second order in $\omega $, $V$ and $T$ for the
lesser and greater self-energies. To this end, it is enough to calculate the
self-energies to order $\widetilde{U}^{2}$, because higher order terms
contribute to higher order in $\omega $, $V$, $T$. The result is given in
terms of renormalized parameters, which in turn can be determined directly
from NRG \cite{re1,re2} or from thermodynamic quantities at equilibrium, for
which accurate (NRG) \cite{bulla} or exact (Bethe ansatz) \cite%
{andr,tsve,bet,pedro} techniques can be applied.

The resulting $\widetilde{\Sigma }^{<}(\omega )$ (obtained by numerical
integration of the diagrammatic expression) is calculated for several values
of $V$ and $T$ and different coupling to the leads and compared with the
analytical expression and in particular to the Ng approximation [Eq. (\ref%
{ngsl})] widely used in different contexts.\cite%
{pz,ser,zz,fer,dong,rap,bal,ding,lim,zhao} While the Ng approximation is
inaccurate and presents artificial jumps at $T=0$ for $|\omega |\lesssim eV$%
, it turns out to be a good approximation in the whole frequency range for $%
k_{B}T\geq eV/2$ for symmetric coupling to the leads or $k_{B}T\geq eV/4$
for the asymmetric cases studied here.

We have also shown that RPT conserves the current to terms of order $(eV / 
\widetilde{\Delta })^{3}$ and discussed the dependence of $\widetilde{\Sigma 
}^{<}(\omega )$ on bias voltage $V$ in terms of Ward identities satisfied by
the analytical approximation.

The analytical results for small energies $\hbar \omega $, $eV$ and $k_{B}T$
compared with the quasiparticle level width $\widetilde{\Delta }$ [Eqs. (\ref%
{sl3}) to (\ref{sret})] can be used to test other approximations for this
tough problem, involving strong correlations out of equilibrium.

The RPT approach to order $\widetilde{U}^{2}$ that we have followed becomes
invalid for $eV>\widetilde{\Delta }$. In particular, it cannot describe the
splitting of the Kondo peak in the spectral density obtained with the
non-crossing approximation,\cite{win,nca2} and observed experimentally in a
three-terminal quantum ring.\cite{letu} This might be corrected by the
inclusion of terms up to fourth order.\cite{fuji}

Concerning physical observables, probably the most studied one in the last
years is the non-equilibrium electric conductance through nanodevices. In the case of
single-level quantum dots for which the impurity Anderson model can be
applied, the lesser and greater quantities can be eliminated from the
expressions of the conductance using conservation of the current.\cite{meir}
The same happens for the energy current and as a consequence also for 
the heat current, as shown in Section \ref{thermal}.
The lesser (or greater) self-energy plays however a role in this
conservation. See Section \ref{cc}. For problems with two levels in which
the couplings to both leads are not proportional, such an elimination is not
possible and the lesser or greater Green functions enter in the expression
for the conductance. An example is the conductance through a benzene molecule
connected to the leads in the \textit{meta} or \textit{ortho} positions, for
which two degenerate levels should be considered (and they couple with
different phases to both leads),\cite{ben}. Other similar systems are molecules with nearly
degenerate even and odd states,\cite{ball},  aromatic molecules or rings
of quantum dots,\cite{rinc2} or two quantum dots connected with different couplings
to two leads.\cite{hart} In these systems, quantum interference plays an
essential role. The case of complete destructive interference is described
by an SU(4) Anderson model,\cite{interf} very similar as the one that
describes carbon nanotubes \cite{jari,ander,buss}, silicon nanowires 
\cite{tetta,see} and more recently a double quantum dot with strong interdot
capacitive coupling, and each QD tunnel-coupled to its own pair of leads,
for certain parameters.\cite{ama,keller,buss2,oks,nishi} The only
difference is that the relevant levels are connected to the leads with
different phases and therefore the conductance is different. Recently RPT
with parameters derived from NRG was applied to this problem for equilibrium
quantities. This approach can be extended to study the interference
phenomena out of equilibrium.

Another observable, directly related to the lesser Green function is the
occupation at the dot, which is given by $\langle n_{d\sigma }\rangle
=-i\int d\omega G_{d\sigma }^{<}(\omega )/(2\pi )$.\cite{none} RPT is not
adequate to calculate this integral because it involves energies far from
the Fermi level.\cite{scali,he1} However, since the difference between $%
\widetilde{\Sigma }^{<}(\omega )$ and the corresponding Ng approximation is
restricted to energies smaller that $eV$ (see Section \ref{nga}), we can
calculate the effect of this approximation on $n=2\langle n_{d\sigma
}\rangle $ using Eq. (\ref{gl}). We find that for the region of parameters
that we have studied, the difference $\Delta n=n -n_{\text{Ng}}$ is very
small, of the order of $10^{-4}z$. This is due to a large compensation of
the regions of positive and negative $\widetilde{\Sigma }_{\text{Ng}%
}^{>}(\omega )-\widetilde{\Sigma }^{<}(\omega )$. In fact using Eqs. (\ref{gl}), 
(\ref{ngsl}) and (\ref{dif}), one realizes that $\Delta n$ is
proportional to $I_L - I_R$ and therefore (from the results of Section \ref{cc}) 
it is of order $z(eV / \widetilde{\Delta })^{4}$.

Nevertheless, one expects that the shortcomings of the Ng approach
would appear in dynamic properties at low frequencies, for which time derivatives enter
the conservation laws and the left and right electric and energy currents become 
different.

We have calculated the effect of the applied bias voltage $V$ on the heat currents between 
any of the leads and the quantum dot. Due to the joule heating, these currents exits 
even at zero temperature for $V \neq 0$. We provide exact expressions to order
$(eV / \widetilde{\Delta })^{3}$ at $T=0$ [Eq. \ref{jq3}]. At finite temperature, 
the current between the dot and one of the leads changes sign as a function of $V$.  

\section*{Acknowledgments}

The author is partially supported by CONICET. This work was sponsored by PIP
112-200801-01821 of CONICET, and PICT 2010-1060 of the ANPCyT, Argentina.

\appendix

\section{Evaluation of the integrals entering the renormalized lesser
self-energy for small energies}

\label{appe}

The integrals entering Eq. (\ref{sl2}) for $\tilde{\Sigma}^{<}(\omega )$
have the form

\begin{eqnarray}
X(\omega ) &=&\int d\epsilon _{2}f(\epsilon _{2}-\mu _{2})Y(\omega ,\epsilon
_{2}),  \label{x} \\
Y(\omega ,\epsilon _{2}) &=&\int d\epsilon _{1}f(\epsilon _{1}-\mu
_{1})f(\omega +\mu _{3}-\epsilon _{1}-\epsilon _{2}).  \label{y}
\end{eqnarray}%
Using

\begin{equation}
f(x)f(y)=\frac{f(-y)-f(x)}{\exp \left( \frac{x+y}{k_{B}T}\right) -1},
\label{fxy}
\end{equation}%
for the integrand of Eq. (\ref{y}) with $x=\epsilon _{1}-\mu _{1}$, $%
y=\omega +\mu _{3}-\epsilon _{1}-\epsilon _{2}$, since $\zeta =x+y$ is
independent of $\epsilon _{1}$, $Y(\omega ,\epsilon _{2})$ becomes
proportional to the integral of a difference of Fermi functions. Using

\begin{equation}
\int dx\left[ f(x-\zeta )-f(x)\right] =\zeta ,  \label{idf}
\end{equation}%
one obtains that $Y(\omega ,\epsilon _{2})$ can be written in terms of the
Bose function $b(\omega )$

\begin{eqnarray}
Y(\omega ,\epsilon _{2}) &=&\zeta b(\zeta ),  \notag \\
b(\zeta ) &=&\frac{1}{\exp \left( \frac{\zeta }{k_{B}T}\right) -1}  \notag \\
\zeta &=&\omega +\mu _{3}-\mu _{1}-\epsilon _{2}.  \label{y2}
\end{eqnarray}%
With the change of variable $v=\epsilon _{2}-\mu _{2}$, replacing Eq. (\ref%
{y2}) in Eq. (\ref{x}) one has

\begin{eqnarray}
X(\omega ) &=&\int dv(a-v)f(v)b(a-v),  \notag \\
a &=&\omega +\mu _{3}-\mu _{1}-\mu _{2}.  \label{x2}
\end{eqnarray}%
Using

\begin{equation}
f(v)b(a-v)=-f(a)\left[ f(v)+b(-a+v)\right] ,  \label{fb}
\end{equation}%
one can write

\begin{eqnarray}
X(\omega ) &=&f(a)\tilde{X}(\omega ),  \label{x3} \\
\tilde{X}(\omega ) &=&-\int dv(a-v)\left[ f(v)+b(-a+v)\right]  \notag \\
&=&\tilde{X}_{1}(\omega )+\tilde{X}_{2}(\omega ),  \label{xt}
\end{eqnarray}%
with

\begin{eqnarray}
\tilde{X}_{1}(\omega ) &=&\int dv(v-a)\left[ f(v)-f(v-a)\right] ,
\label{xt1} \\
\tilde{X}_{2}(\omega ) &=&\int ydy\left[ f(y)+b(y)\right] =(k_{B}T)^{2}\int
dx\frac{x}{\sinh (x)}  \notag \\
&=&\frac{\pi ^{2}}{2}(k_{B}T)^{2}.  \label{xt2}
\end{eqnarray}%
Above, the changes of variable $y=v-a$, $x=y/(k_{B}T)$ were used.

Using instead $y=v-a/2$, $\tilde{X}_{1}$ becomes

\begin{eqnarray}
\tilde{X}_{1}(\omega ) &=&\int ydy\left[ f(y+a/2)-f(y-a/2)\right]  \notag \\
&&-\int dv\left[ f(v)-f(v-a)\right] a/2.  \label{xt1b}
\end{eqnarray}%
The first integral vanishes, since the integrand is odd [as can be checked
using Eq. (\ref{one})]. Using Eq. (\ref{idf}) the second integral gives $%
\tilde{X}_{1}(\omega )=a^{2}/2$. Replacing this and Eq. (\ref{xt2}) in Eq. (%
\ref{x3}) we finally obtain

\begin{equation}
X(\omega )=\frac{f(a)}{2}\left[ a^{2}+(\pi k_{B}T)^{2}\right] .  \label{x4}
\end{equation}


\begin{thebibliography}{99}
\bibitem{note2} The failure of the argument in the reply \cite{reply} in its
claiming that the Ward identity is not satisfied is due to the fact that it
is based on an inappropriate expansion of $\Sigma ^{<}(\omega ,V)$ for $T=0$
around the singular point $\omega =V=0$. To illustrate the point, let us
consider the function [similar to Eq. (\ref{sl3}) at $T=0$] $%
F(\omega,V)=\theta (-\omega +V)(\omega -V)^{2}+\theta (-\omega -V)(\omega
+V)^{2}$, where $\theta (x)$ is the step function. Expanding this function
to total second order around the origin gives $\tilde{F}(\omega ,V)\simeq
2(\omega ^{2}+V^{2})\theta (0)$, which is obviously very different from $F$
out of the origin. In particular even for a tiny $V>|\omega |$, $F=(\omega
+V)^{2}$. According to Ref. \onlinecite{reply}, $F$ cannot have a term in $%
\omega V$ for small $\omega $ and $V$ because $\tilde{F}$ does not have it.
\bibitem{note} The precise values of the current depend on the exact value
of $\widetilde{U}/(\pi \widetilde{\Delta })$, but this does not modify our
conclusions.
\end{thebibliography}
\end{document}